\documentclass[useAMS,usenatbib,usegraphicx]{mn2e}
\usepackage{times}

\newcommand{\kep}{{\em Kepler}}

\newcommand{\kepmi}{{\em Kepler Mission}}

\newcommand{\teff}{\ensuremath{T_{\rm{eff}}}}
\newcommand{\logg}{\ensuremath{\log g}}

\newcommand{\logZ}{\ensuremath{\log [\rm Z]}}
\newcommand{\lheh}{\ensuremath{\log \left(\rm N_{\mathrm{He}}/\rm N_{\mathrm{H}}\right)}}

\newcommand{\lsim}{\raisebox{-1ex}{$\stackrel{{\displaystyle<}}{\sim}$}}
\newcommand{\gsim}{\raisebox{-1ex}{$\stackrel{{\displaystyle>}}{\sim}$}}

\newcommand{\msun}{${\mathrm{M}}_{\odot}$} 
\newcommand{\rsun}{${\mathrm{R}}_{\odot}$}

\title[Orbital properties of an unusually low-mass sdB star in a close binary 
with a white dwarf]
{Orbital properties of an unusually low-mass sdB star in a close binary system
with a white dwarf}

\author[R.~Silvotti et al.]
       {R.~Silvotti$^1$\thanks{E-mail: silvotti@oato.inaf.it},
        R.~H.~\O stensen$^2$,
	S.~Bloemen$^2$,
        J.~H.~Telting$^{3}$,
	U.~Heber$^{4}$,
        R.~Oreiro$^{5}$,\newauthor
        M.~D.~Reed$^{6}$,
	L.~E.~Farris$^{6}$,
	S.~J.~O'Toole$^{7}$,
	L.~Lanteri$^1$,
	P.~Degroote$^2$,
	H.~Hu$^{8}$,\newauthor
	A.~S.~Baran$^{6,9}$,
	J.~J.~Hermes$^{10}$,
	L.~G.~Althaus$^{11}$,
	T.~R.~Marsh$^{12}$,
	S.~Charpinet$^{13}$,
	J.~Li$^{14}$,\newauthor
	R.~L.~Morris$^{14}$,
	D.~T.~Sanderfer$^{14}$\\\\
        $^1$ INAF-Osservatorio Astrofisico di Torino,
             Strada dell'Osservatorio 20, 10025 Pino Torinese, Italy\\
        $^2$ Instituut voor Sterrenkunde, KU Leuven, 
	     Celestijnenlaan 200D, 3001 Leuven, Belgium\\
        $^3$ Nordic Optical Telescope, Apartado 474, 
	     38700 Santa Cruz de La Palma, Spain\\
        $^4$ Dr. Karl Remeis-Observatory \& ECAP, Astronomical Inst.,
             FAU Erlangen-Nuremberg, Sternwartstr.~7, 96049 Bamberg, Germany\\
        $^5$ Instituto de Astrof\'isica de Andaluc\'ia,
             Glorieta de la Astronom\'ia s/n, 18008 Granada, Spain\\
        $^6$ Department of Physics, Astronomy and Materials Science,
             Missouri State University, Springfield, MO 65897, USA\\
        $^7$ Anglo-Australian Observatory, PO Box 296, Epping, NSW 1710,
	     Australia\\
	$^8$ Institute of Astronomy, The Observatories, Madingley Road, 
	     Cambridge CB3 0HA, UK\\
	$^9$ Uniwersytet Pedagogiczny, Obserwatorium na Suhorze, 
	     ul. Podchor\c{a}\.{z}ych 2, 30-084 Krak\'ow, Polska\\
	$^{10}$ Department of Astronomy, University of Texas at Austin, 
	        Austin, TX 78712, USA\\
	$^{11}$ Facultad de Ciencias Astron\'omicas y Geof\'isicas, 
	        Universidad Nacional de La Plata, La Plata 1900, Argentina\\
	$^{12}$ Department of Physics, University of Warwick, Coventry CV4 7AL,
	        UK\\
	$^{13}$ Laboratoire d'Astrophysique de Toulouse-Tarbes, Universit\'e 
	        de Toulouse, 14 avenue Edouard Belin, Toulouse 31400, France\\
	$^{14}$ SETI Institute/NASA Ames Research Center, Moffett Field, 
	        CA 94035, USA\\
}

\begin{document}

\date{Released 2012 Xxxxx XX}
\pagerange{\pageref{firstpage}--\pageref{lastpage}} \pubyear{2012}
\maketitle
\label{firstpage}

\begin{abstract}
We have used 605 days of photometric data from the \kep\ spacecraft to study
KIC~6614501, a close binary system with an orbital period of 
0.15749747(25) days (3.779939 hours), 
that consists of a low-mass subdwarf B (sdB) star and a white dwarf.
As seen in many other similar systems, the gravitational field of 
the white dwarf produces an ellipsoidal deformation of the sdB which appears 
in the light curve as a modulation at two times the orbital frequency.
The ellipsoidal deformation of the sdB implies that the system has a maximum
inclination of $\sim$40 degrees,
with $i$~$\approx$~$20^{\circ}$ being the most likely.
The orbital radial velocity of the sdB star is high enough to produce a Doppler
beaming effect with an amplitude of 432~$\pm$~5~ppm,
clearly visible in the folded light curve.
The photometric amplitude that we obtain, 
$K_1=85.8 ~ \rm km/s$, is $\sim$12 per cent less than the spectroscopic RV 
amplitude of $97.2 \pm 2.0 ~ \rm km/s$.
The discrepancy is due to the photometric contamination from a close object at 
about 5 arcsec North West of KIC~6614501, which is difficult to remove.
The atmospheric parameters of the sdB star, $\teff=23~700 \pm 500~ \rm K$ and
$\logg=5.70 \pm 0.10$, imply that it is a rare object below the Extreme 
Horizontal Branch (EHB), similar to HD~188112 \citep{heber03}.
The comparison with different evolutionary tracks suggests a mass between
$\sim$0.18 and $\sim$0.25 \msun, too low to sustain core helium burning.
If the mass was close to 0.18-0.19 \msun, the star could be already on the 
final He-core WD cooling track.
A higher mass, up to $\sim$0.25 \msun, 
would be compatible with a He-core WD progenitor undergoing a cooling phase
in a H-shell flash loop.
A third possibility, with a mass between $\sim$0.32 and $\sim$0.40 \msun,
can not be excluded and would imply that the sdB is a ``normal'' (but with an
unusually low mass) EHB star burning He in its core.
In all these different scenarios the system is expected to merge in less than 
3.1 Gyr due to gravitational wave radiation.
\end{abstract}

\begin{keywords}
binaries: close --
hot subdwarfs --
white dwarfs --
stars: individual: KIC~6614501
\end{keywords}

\section{Introduction}

Hot subdwarf stars (sdBs and sdOs) are found in all galactic stellar 
populations and they are the main source of the UV-upturn phenomenon in 
early-type galaxies \citep{greggio90,brown97}.
Subdwarf B stars in particular are post red giant branch (RGB) stars with thin 
(\lsim 0.01 \msun) inert hydrogen envelopes.
Most of them have masses close to 0.5 \msun, with a peak near 0.47, and
are core helium-burning objects on the extreme horizontal branch (EHB).
If the mass of an sdB star is below $\sim$ 0.3 \msun, the He-core ignition 
never took place.
These rare stars are the progenitors of He-core white dwarfs.
A recent extensive review on sdB/sdO stars is given by \citet{heber09}.

In order to reach the high temperatures and surface gravities typical of sdB 
stars, their progenitors must have lost almost the entire hydrogen envelope 
near the tip of the red giant branch (RGB).
This may happen in a close binary after a common-envelope (CE) phase, when the 
orbital angular momentum is transferred to the red giant envelope spinning up 
the envelope until enough energy has been accumulated to eject it.
Indeed, about half of sdBs reside in close binary systems with white 
dwarfs (WDs) or low-mass main-sequence stars \citep{maxted01,napiwotzki04}.
The various formation channels of sdB stars have been studied by 
\citet{han02,han03}.

Presently, more than one hundred close binaries with an sdB component are 
known \citep{geier11a, geier11b,copperwheat11}, and an important fraction must 
be composed by sdB+WD binaries.
These systems are potential progenitors of type Ia supernovae (SN), provided 
that the total mass exceeds the Chandrasekhar limit and that the initial 
separation is small enough to merge in a Hubble time due to gravitational wave 
radiation.
The first candidate found to be a type Ia SN progenitor is
KPD~1930+2752 \citep{maxted00,geier07}, with a total mass of 1.47 \msun\ 
when we assume a canonical value of 0.5 \msun\ for the sdB component.
Another interesting sdB+WD system, KPD~1946+4340, was observed by the \kep\ 
spacecraft and studied in great detail \citep{bloemen11}.
Thanks to its brithness (V=14.28) and high inclination (87.1$^{\circ}$),
KPD~1946+4340 shows very clean primary and secondary eclipses and a very 
accurate light curve modelling was possible, including not only ellipsoidal 
deformation and Doppler beaming, but also WD reflection and gravitational 
lensing.

The target discussed in this article, KIC~6614501 (alias 2MASS 
J19365001+4201436), is another such sdB+WD system.
It was selected among a list of sdB pulsator candidates inside 
the \kep\ field of view, with the main goal of performing detailed 
asteroseismic studies of these stars.
During the {\it survey phase} \citep{ostensen10b,ostensen11},
it was observed by the \kep\ space telescope \citep{borucki10}
in the framework of the \kep\ Asteroseismic Science Consortium 
\citep{gilliland10}.
The first results after 1 month of observation are reported by 
\citet{ostensen11}: no pulsations were detected and the atmospheric parameters
were determined from spectroscopy at the 4.2~m William Herschel Telescope.
A low frequency modulation plus its 1st harmonic in the \kep\ light curve was 
interpreted as a binary signature, suggesting a sdB+WD system because no 
red excess was seen in the spectrum of the star.
\citet{ostensen11} also noted the peculiar position of KIC~6614501 in the
\teff/\logg\ plane, below the EHB, implying a low-mass post-RGB object.
A mass of about 0.24 \msun\ was suggested by comparing its position with the 
evolutionary tracks of \citet{driebe98}, implying that KIC~6614501 could
evolve into a He-core white dwarf.

This particular configuration may occur when the envelope of the red giant 
is ejected before the core has attained sufficient mass to ignite helium.
In this case, the RG core may cross the region of the EHB stars near its low
temperature - low gravity edge, before evolving to the WD cooling track.
These post-RGB low-mass objects are rare and only recently their number has
increased significantly thanks in particular to the results of the ELM=Extreme 
Low Mass (WD) survey \citep{brown10}, the \kep\ mission, and the WASP 
(Wide Angle Search for Planets) survey (e.g. \citealt{maxted11}).
The ELM survey has found many new double degenerate systems, increasing the 
number of those previously discovered by the SPY survey \citep{koester09}. 
A spectacular case is J1741+6526, composed by a ELM WD primary of 0.16 \msun\
and an unseen companion with a minimum mass of 1.1 \msun:
with an orbital period of 1.47 hours, it should merge in less than 170 Myr. 
Assuming an inclination of 60 degrees, the total mass would be 1.7 \msun\ 
\citep{brown12}.

However, most ELM WDs are concentrated at effective temperatures below 
15~000~K.
At higher temperatures, the number of known ELM WDs (or ELM WD precursors) 
is very small, with only a few candidates.
KIC~6614501 is the hottest candidate in a region of the \teff/\logg\ plane 
almost completely empty and crucial to understand the evolution of the stars
that have a mass close to the limit for He-core burning.
The system most similar to KIC~6614501 is HD~188112, with a $\sim$0.24 \msun\ 
sdB primary and a massive ($>$0.73 \msun) unseen companion \citep{heber03}.
If the companion was a C/O WD with a mass larger than 0.9 \msun, HD~188112
would be the precursor of a subluminous SN Ia.

In the next sections we will perform a detailed study of the properties of
KIC~6614501, based on \kep\ data plus spectroscopic follow-up.
In the last section of this article, our results are placed in 
the context of the ELM white dwarfs.

\section{Kepler light curve and ephemeris}

KIC~6614501 was observed by \kep\ in short cadence (SC, 59 s sampling) 
during the following monthly runs (``Q'' stands for quarter): Q3.3, Q5.1, Q5.2,
Q5.3, Q6.1, Q6.2, Q6.3, Q8.1, Q8.2, Q8.3, Q9.1, Q9.2, Q9.3, Q10.1, Q10.2 and 
Q10.3 for a total duration of 458.9 days.
Moreover it was observed in long cadence (LC, 29.4 min sampling) during Q3.1, 
Q3.2 and Q7 (146.4 days in total).
The data were downloaded from the \kep\ Asteroseismic Science Operations 
Center (KASOC) website~\footnote{http://kasoc.phys.au.dk/kasoc/}.
The data files contain Barycentric Julian Dates (BJD), raw fluxes with 
errors, and corrected fluxes with errors.
The corrected fluxes are computed by estimating and subtracting the 
contamination by close objects, based on their spectral energy distribution.
However, the colour-based atmospheric parameters (\teff\ and \logg) of the 
stars in the \kep\ database are 
not 
accurate 
for stars with peculiar properties such as very hot stars.
For this reason we decided to use raw fluxes.
This choice may affect the relative amplitude of the orbital modulation and its
1st harmonic: a discussion of this aspect is given in section 4.1.

For each run the fluxes were corrected for long-term trends using a cubic 
spline interpolation or a straight line in a few cases.
We removed outliers ($>$4 $\times$ standard deviation), and the fluxes were 
divided by their mean value in order to normalize all runs to an average value 
of 1.

A first Fourier analysis up to the Nyquist frequency was performed using only 
the short cadence data to check for any possible presence of high-frequency 
oscillations.
Excluding some known artifacts, only a few peaks were found with an amplitude
just beyond the threshold which we fixed at 4 times the mean value of the 
amplitude spectrum (4$\sigma$). The highest peak is at 579.319 d$^{-1}$
(670.508 $\mu$Hz), with an amplitude of 28.9 ppm, corresponding to 4.3$\sigma$.
With more data from \kep\ it will be possible in the future to confirm or not
the presence of high-frequency oscillations in KIC~6614501.

The low-frequency analysis was done using all the 676~744 data points (SC + 
LC), keeping their original sampling time.
We verified that the smearing effect introduced by LC data was smaller than the
amplitude uncertainty.
At low frequency, the Fourier amplitude spectrum shows two peaks that 
correspond to the orbital period and its first harmonic.
The latter is the result of the ellipsoidal deformation of the sdB star
due to the gravitational field produced by the companion.
Less obvious is the correct interpretation of the orbital modulation, 
which requires also the phase information (see section 4).
The light curve and the amplitude spectrum at low frequency are shown in 
Fig.~1. 

\begin{figure}
\centering
\includegraphics[width=8.8cm, angle=0]{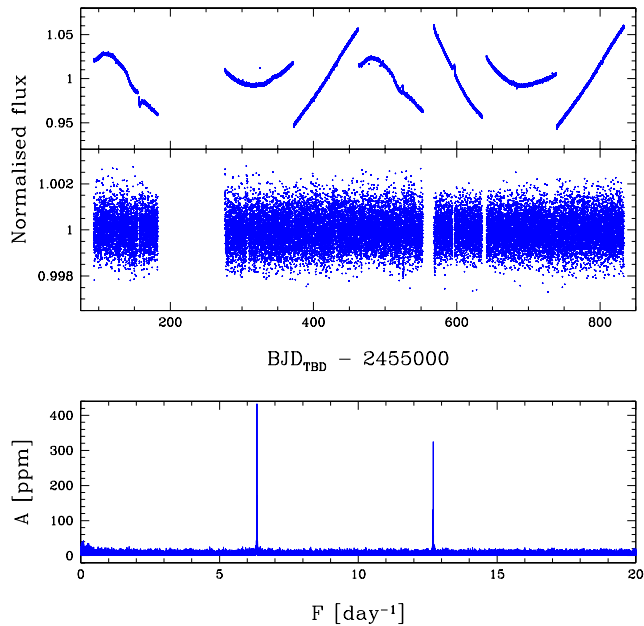}
\caption{Top: the \kep\ raw light curve after normalisation of each 
segment (upper panel) and after cubic spline (or straight line) correction
(middle panel).
Bottom: amplitude spectrum at low frequency showing the orbital period and its 
first harmonic.
Only in the two upper panels, in order to reduce the size of the figures,
we used a light curve with a reduced number of points by binning to 29.4 min 
all the short cadence data (while in the lower panel and in all our analysis
the data were used with their original sampling time).
}
\label{fig:lc_dft}
\end{figure}

From fitting the orbital period to the \kep\ light curve, we obtain the
following ephemeris:

\vspace{2mm}

\noindent
%
BJD$_{\rm TDB}$ = 2455093.33296(31) + 0.15749747(25) $E$
%
\vspace{2mm}

\noindent
which gives the times when the sdB is closest to the Earth 
(i.e. phase 0 in Fig. 4).
The reference epoch 
corresponds to the first time at phase 0 of the \kep\ data.

\section{Spectroscopy}

On the nights of 12 and 15 August 2010 we obtained 22 spectra of KIC~6614501
with the Kitt Peak National Observatory (KPNO) 4-m Mayall telescope and its 
RC-Spec/F3KB spectrograph.
The spectra cover the H$_\beta$--H$_{\eta}$ region with a dispersion of 
0.45~\AA/pix and a resolution of 1.6~\AA.
The exposure time was 600 seconds, yielding a median S/N=38 per pixel.

Similarly, on the night of 27 June 2010, we obtained 10 spectra at the 2.5-m 
Isaac Newton Telescope (INT) with the IDS235+EEV10 spectrograph, with a
resolution of 3.0~\AA, exposure time of 1600 seconds, and median S/N=42
when resampled to the KPNO 4m dispersion.

The data were homogeneously reduced and analysed. 
Standard reduction steps within IRAF~\footnote{Image Reduction and Analysis 
Facility, written and supported by the National Optical Astronomy Observatories
(NOAO) in Tucson, Arizona ({\it http://iraf.noao.edu/}).}
were: bias subtraction, removal of pixel-to-pixel sensitivity variations, 
optimal spectral extraction, and wavelength calibration based on arc lamp 
spectra.
The target spectra and mid-exposure times were shifted to the barycentric frame
of the solar system. 
The spectra were normalised to place the continuum at unity by comparing with a
model spectrum for a star with similar physical parameters as we find for the 
target (see section 3.2).
 
\subsection{Spectroscopic radial velocities}

Radial velocities (RVs) were derived with the FXCOR package in IRAF using the
H$_\beta$, H$_\gamma$, H$_\zeta$ and H$_\eta$ lines.
All the RV measurements are reported in Table~1.
%
Fits assuming a sinusoidal orbital velocity curve confirm the orbital period as
0.157497(5)~d, fully consistent with the photometric orbital period found from 
the \kep\ data.
The RV amplitude that we find is $97.2 \pm 2.0$ km/s.
Due to the relatively long exposure times of the INT data, the measured 
velocities are smeared-out to result in an RV amplitude that is lower than the 
fit by about 2 km/s.
For the KPNO data the smearing effects due to the exposure time are negligible.
When fitting the data of each observatory with the period and phase fixed from 
the all-data fit, we found that the RV amplitude is consistent.
The system velocity is $-9.9(2.2)$ km/s for the KPNO data, and $0.5(3.2)$ km/s 
for the INT data.
This zero point offset also shows up in the position of the
interstellar CaII 3993~\AA\ line in the mean spectrum of each telescope.
Although the measurements of the position of this line are not accurate 
enough to calibrate this offset, it is clear that the offset is 
site dependent and does not imply the presence of a third body.
Our best fit is shown in Fig.~2 and the fit parameters are listed in Table~2.
The orbital period was fixed to the value determined from \kep\ photometry.

After the orbital fits, the spectra were shifted to remove the orbital motion, 
before being co-added to obtain high S/N spectra (S/N$\sim$150) with minimal 
orbital line broadening, for both observatories.
These spectra were used to derive the atmospheric parameters of the star.

\begin{table} \centering
\caption[]{RV measurements.}
\begin{tabular}{crr}
\hline
BJD & v (km/s) &  error (km/s)\\
\hline
2455375.481497 &   13.5~~ & 25.0~~~~~~ \\
2455375.500704 &  -66.8~~ & 15.9~~~~~~ \\
2455375.520005 &  -94.0~~ &  6.6~~~~~~ \\
2455375.541801 &  -48.6~~ & 11.7~~~~~~ \\
2455375.561105 &  -14.8~~ & 10.4~~~~~~ \\
2455375.579773 &   60.7~~ &  5.3~~~~~~ \\
2455375.601982 &   76.1~~ &  8.1~~~~~~ \\
2455375.624646 &   75.2~~ &  5.7~~~~~~ \\
2455375.649749 &  -33.6~~ &  8.0~~~~~~ \\
2455375.675079 & -102.8~~ &  8.0~~~~~~ \\
2455420.659287 &   71.5~~ & 11.6~~~~~~ \\
2455420.671652 &   35.3~~ & 11.3~~~~~~ \\
2455420.678830 &   22.7~~ & 10.1~~~~~~ \\
2455420.685997 &    2.8~~ &  6.6~~~~~~ \\
2455420.693178 &  -45.0~~ &  7.5~~~~~~ \\
2455420.705115 &  -70.7~~ &  6.8~~~~~~ \\
2455420.713657 & -104.5~~ &  5.0~~~~~~ \\
2455420.720826 &  -98.5~~ &  9.4~~~~~~ \\
2455420.759432 &  -29.5~~ &  9.2~~~~~~ \\
2455420.766993 &   -1.5~~ &  9.7~~~~~~ \\
2455420.774178 &   23.9~~ &  6.6~~~~~~ \\
2455420.781347 &   45.6~~ &  5.5~~~~~~ \\
2455420.788526 &   71.8~~ & 11.9~~~~~~ \\
2455423.635478 &   74.3~~ &  7.9~~~~~~ \\
2455423.642873 &   80.7~~ & 10.8~~~~~~ \\
2455423.650043 &   87.8~~ &  9.4~~~~~~ \\
2455423.657195 &   67.2~~ &  5.0~~~~~~ \\
2455423.664441 &   55.8~~ &  8.2~~~~~~ \\
2455423.716534 & -102.8~~ &  8.7~~~~~~ \\
2455423.724074 &  -99.8~~ & 12.9~~~~~~ \\
2455423.731241 &  -85.5~~ & 30.4~~~~~~ \\
2455423.738429 &  -82.0~~ & 40.5~~~~~~ \\
\hline
\end{tabular}
\end{table}

\begin{figure}
\centering
\includegraphics[width=6.4cm, angle=270]{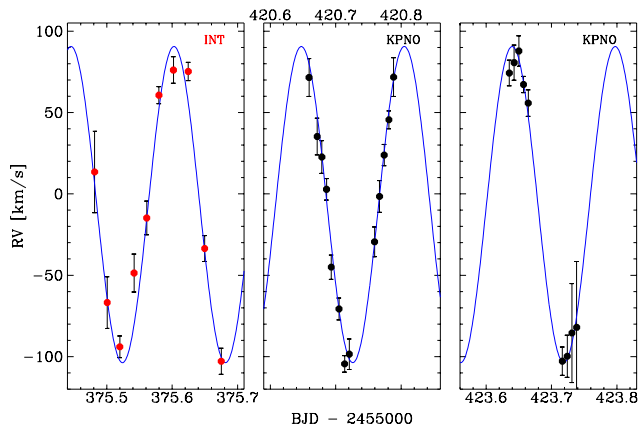}
\caption{The radial velocity curve of KIC~6614501 using the orbital period as 
determined from \kep\ photometry. See the text for more details.}
\label{fig:RVs}
\end{figure}

\begin{table} \centering
\caption[]{Orbital fit parameters using the orbital period from 
\kep\ photometry.}
\begin{tabular}{lc}
\hline
Offset (km/s)          & -6.5 $\pm$ 1.5 \\
Amplitude (km/s)       & 97.2 $\pm$ 2.0 \\
Period (day)           & 0.15749747 \\
Phase~0 (BJD--2455000) & 375.56366 $\pm$ 0.00054 \\
$\chi^2_{red}$         & 1.58 \\
\hline
\end{tabular}
\end{table}

%
%

\subsection{Improved atmospheric parameters}

The high S/N mean spectra from the INT and KPNO, shown in Fig.~3, were fitted 
to a grid of LTE spectra computed {\it ad hoc} for this star \citep{heber00}.
Since there is no detectable helium lines in the spectra, we fixed the helium 
abundance to $\lheh = -3.0$.
The grids were computed at solar metallicity
and at metallicities reduced by a factor of 10 or 100 relative to solar 
composition. 
From an evaluation of the line strength in the models when convolved with the 
instrumental profile, we infer, from the non-detection of any metal features in
our spectra, that the overall metallicity must be significantly sub-solar, 
at least $\logZ < -1.5$.
But we know that individual elements may be substantially under- or 
overabundant relative to the solar composition, as is typical for the sdB stars
due to the competing effects of gravitational settling and radiative levitation
and to possible radiation-driven wind (see section 5 of \citealt{heber09} and
references therein).
In the end we decided to use the grid computed at 
$\logZ = -2.0$ for the final analysis.
The H$_\beta$ line was kept out of the fit, partly due to a problem with the 
background subtraction in the KPNO data, and partly due to concerns regarding 
NLTE effects in the core, which are not accounted for by our models. 
For the KPNO data set the H$_\eta$ line was also kept out due to a bad CCD 
column which translates into a broad artifact feature in the mean spectrum 
after applying the orbit correction.
Our best atmospheric parameters from INT and KPNO spectra are summarised in 
Table~3.

\begin{figure}
\centering
\includegraphics[width=8.8cm, angle=270]{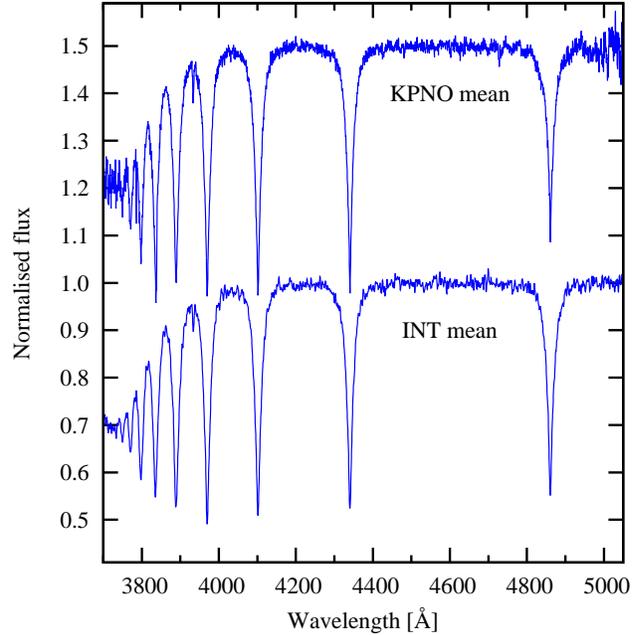}
\caption{The high S/N mean spectra from the INT and KPNO.}
\label{fig:spectra}
\end{figure}

\begin{table} \centering
\caption[]{Best atmospheric parameters from INT and KPNO spectra.}
\begin{tabular}{lcc}
\hline
Data & $T_{\rm eff}$ & log($g$) \\
\hline
INT  & 24~084[65] & 5.732[11] \\
KPNO & 23~332[80] & 5.653[12] \\
\hline
Adopted & 23~700 $\pm $500 & 5.70 $\pm$ 0.1 \\
\hline
\end{tabular}
\end{table}

\section{Light curve analysis}

The light curve of KIC~6614501 can be fit using two components:
the ellipsoidal deformation of the sdB star and the Doppler beaming.
Our best fit to the data with two sinusoidal components is shown in Fig.~4.
We obtain $\chi^2_{red}$=1.17.
The amplitudes of the orbital beaming frequency and of the ellipsoidal 
modulation at two times the orbital frequency are
$432.4 \pm 5.3$ and $324.8 \pm 5.3$ ppm respectively.
A third component due to the reflection effect by the illuminated hemisphere
of the companion is too faint to be seen in the light curve and can be 
neglected.
This component would have the maximum at phase 0.5 and is not seen in the
residuals (bottom panel of Fig.~4).
The lack of an observable reflection effect suggests that the companion must be
a white dwarf. With such a short orbital period, an M-dwarf companion would be 
easily seen (see e.g. the folded light curves of KIC~1868650, KIC~9472174
\citep{ostensen10a,ostensen10b}, KIC~2991403, KIC~11179657 \citep{kawaler10} 
and KIC~7335517 \citep{ostensen11}).
%

\begin{figure}
\centering
\includegraphics[width=8.8cm]{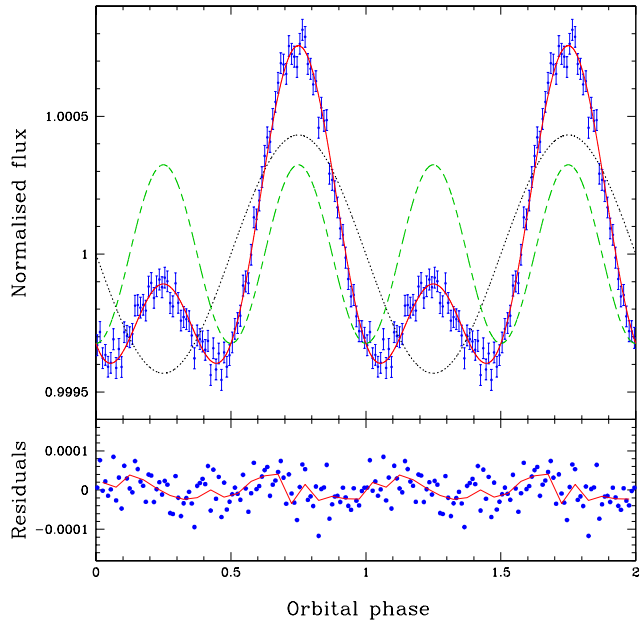}
\caption{Phase-folded light curve of KIC~6614501 (data points grouped in 100 
phase intervals) and our best fit, obtained considering two components:
an orbital modulation that accounts for the beaming effect (black dotted) 
and an ellipsoidal modulation at twice the orbital frequency (green dashed).
The bottom panel shows the residuals grouped in 100 (blue dots) and 20 
(red line) phase intervals.
Colors are available only in the electronic version.}
\label{fig:phas}
\end{figure}

\subsection{Doppler beaming and photometric RVs}

The asymmetry in the ellipsoidal modulation of KIC~6614501 is caused by 
Doppler beaming \citep{hills74, loeb03}.

Doppler beaming, also called Doppler boosting, was first detected by 
\citet{maxted00} in the sdB+WD binary KPD~1930+2752.
Recent detections of Doppler beaming are reported
by \citet{shporer10} and \citet{vennes11}, using data from the ground,
and by \citet{mazeh10}, who have measured the Doppler beaming caused 
by a substellar companion in CoRoT data.
In the context of $Kepler$, Doppler beaming was discussed by 
\citet{loeb03} and \citet{zucker07}.
Thanks to \kep's sensitivity, \citet{vankerkwijk10} have obtained the 
first photometric radial velocities (RVs) for the two A/B+WD systems
KOI~74 and KOI~81.
For KOI~74 the photometric RVs were confirmed a-posteriori by 
spectroscopic measurements \citep{bloemen12}.
A good agreement between photometric and spectroscopic RVs was found also
for the two sdB+WD binaries KPD~1946+4340 \citep{bloemen11} and 
KIC~11558725 \citep{telting12}.

When a source moves at a radial nonrelativistic speed $v_r$ with respect to an 
observer, the observed flux $F_{\lambda}$ is given by:

\begin{center}
\begin{equation}
F_{\lambda} = F_{0 \lambda} \left( 1 - B_{\lambda} \frac{v_r}{c}\right)
\label{eq:B}
\end{equation}
\end{center}

\noindent
where $F_{0 \lambda}$ is the emitted flux and $c$ is the light speed.\\
$B_{\lambda} = B_g + B_{s \lambda} = 3 + (2 +\rm d ln \it F_{\lambda} / \rm d 
ln \lambda) = 5 + \rm d ln \it F_{\lambda} / \rm d ln \lambda$ 
is the beaming factor \citep{loeb03, bloemen11},
which incorporates two different terms:
a geometrical term $B_g=3$, not depending on wavelength,
that includes a +1 contribution (enhanced photon arrival rate of an 
approaching source) and a +2 contribution (solid angle geometrical 
aberration).
And the Doppler shift $B_{s \lambda}$ that can either increase 
or decrease the flux of an approaching source, depending on its spectral 
energy distribution (SED) and on the instrumental bandpass.
In our case, for a hot star, the optical flux is on the Rayleigh-Jeans tail of 
the distribution and therefore, when the star is approaching, the blue-shift
pushes part of the flux outside the visibility range giving a negative 
contribution to the beaming factor.

When equation (1) 
is used in broadband photometry, we need to 
compute a weighted mean of the beaming factor that takes into account the
star's SED and the bandpass, which in turn depends on the instrumental response
function ($\epsilon_{i \lambda}$), atmospheric transmission 
($\epsilon_{a \lambda}$) and interstellar medium transmission
($\epsilon_{im \lambda}$):

\begin{center}
\begin{equation}
\langle B \rangle = \frac{\int \epsilon_{i \lambda} ~ \epsilon_{a \lambda} ~ 
\epsilon_{im \lambda} ~ \lambda ~ F_{\lambda} ~ B_{\lambda} ~ d\lambda}
{\int \epsilon_{i \lambda} ~ \epsilon_{a \lambda} ~ \epsilon_{im \lambda} ~ 
\lambda ~ F_{\lambda} ~ d\lambda}
\label{eq:avB}
\end{equation}
\end{center}

\noindent
In our case $\epsilon_{i \lambda}$ is the response function of the \kep\ 
bandpass, $\epsilon_{a \lambda} \equiv 1$ 
at any $\lambda$, and 
$\epsilon_{im \lambda}$ is the interstellar medium transmission corresponding
to a reddening $\rm E(B-V)=0.06 \pm 0.03~mag$ (95\% confidence interval).
This value was obtained by collecting multicolour photometry from the Carlsberg
Meridian Catalog (SDSS r' band, ViZieR catalog I/304), the \kep\ Input Catalog 
(SDSS g,r,i and z bands, ViZieR catalog V/133)  and the UVEX survey 
\citep[UVEX u, g, r, i and H$\alpha$ bands, ][]{groot09}, and by following the
method described in \citet{degroote11}, but applied to a grid of sdB spectral 
energy distributions. 

In order to compute the beaming factor and its uncertainty from equation (2), 
we have used a grid of fully metal line-blanketed LTE atmosphere models for 
sdB stars \citep{heber00}
with \teff\ ranging from 21~000 to 25~000~K, \logg\ between 5.1 and 5.9,
metallicity between solar and 1/100 solar and fixed helium abundance
$\lheh =-3$.
The beaming factor is found to be $\langle B \rangle = 1.51 \pm 0.02$
(1.52 without considering reddening).
This value corresponds to a metallicity of 1/100 solar.
The uncertainty incorporates any value of the metallicity between 1/100 solar 
and solar (metallicity is the main source of error), and an uncertainty
of $\pm500$~K and $\pm0.1$~dex on the sdB's effective temperature and gravity.

Once the beaming factor is computed, in order to calculate the photometric 
radial velocity amplitude, equation (1) can be written as:

\begin{equation}
K = \frac{c ~ A_B}{\langle B \rangle} 
\label{eq:RV}
\end{equation}

\noindent
where $A_B$ is the beaming amplitude.
From equation (3) we obtain a photometric RV amplitude
$K_1 = 85.8 ~ \rm km/s $, 11.7\%
less than the spectroscopic RV amplitude of $97.2 \pm 2.0 ~ \rm km/s $.
The reason of this discrepancy is the photometric contamination from a close 
object, at approximately 5.0 arcsec North West of KIC~6614501, while
the \kep's pixel size is 3.98 arcsec (see Fig.~5).
This flux excess dilutes the beaming modulation and decreases its 
amplitude~\footnote{Even using the \kep\ pixel data it is not possible 
to exclude the photons coming from the close source in a reliable way because 
the coordinates of the pixel mask change constantly in time, see caption of 
Fig.~5.}.
%
%
The uncertainty on the photometic RV amplitude is dominated by a systematic 
error given by the photometric contamination, which is difficult to estimate.
Then we have two statistical errors associated with the beaming factor 
($\pm$1.3\%) and to the beaming amplitude ($\pm$1.2\%), 
coming from the fit of the light curve.

\subsection{Amplitude of the ellipsoidal deformation}

From the comparison between photometric and spectroscopic RV amplitude, we have
seen that only 88.3\% 
of the photons attributed to KIC~6614501 actually come from this source.
This means that not only the beaming amplitude but also the ellipsoidal 
modulation amplitude is reduced by the same amount.
Thus we estimate that the amplitude of the ellipsoidal deformation, after
correction for the photometric contamination, is $368.0 \pm 11.7$ ppm.

\begin{figure}
\centering
\begin{minipage}[c]{4.15cm}
\includegraphics[height=4.125cm]{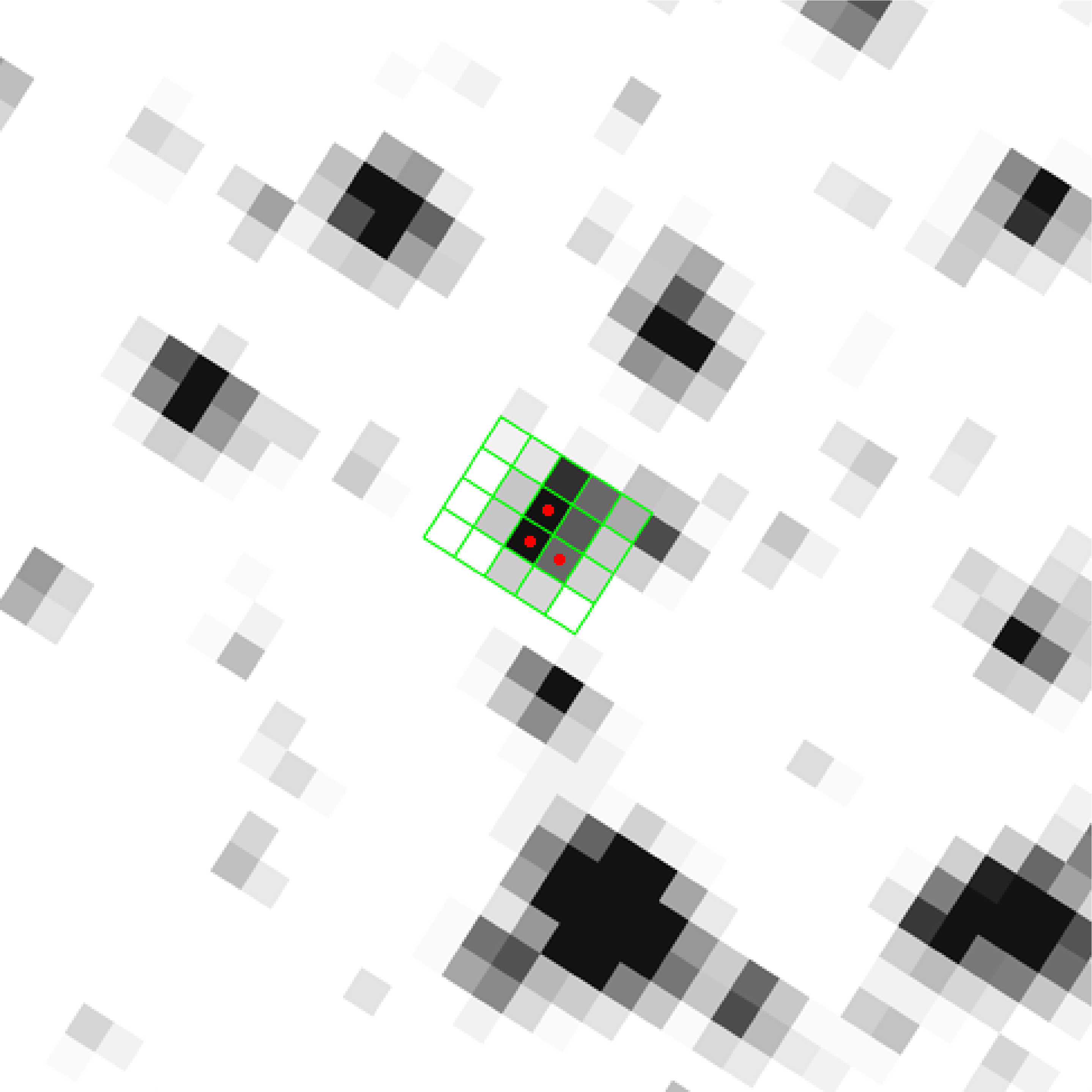}
\end{minipage}
\begin{minipage}[c]{4.15cm}
\includegraphics[height=4.125cm]{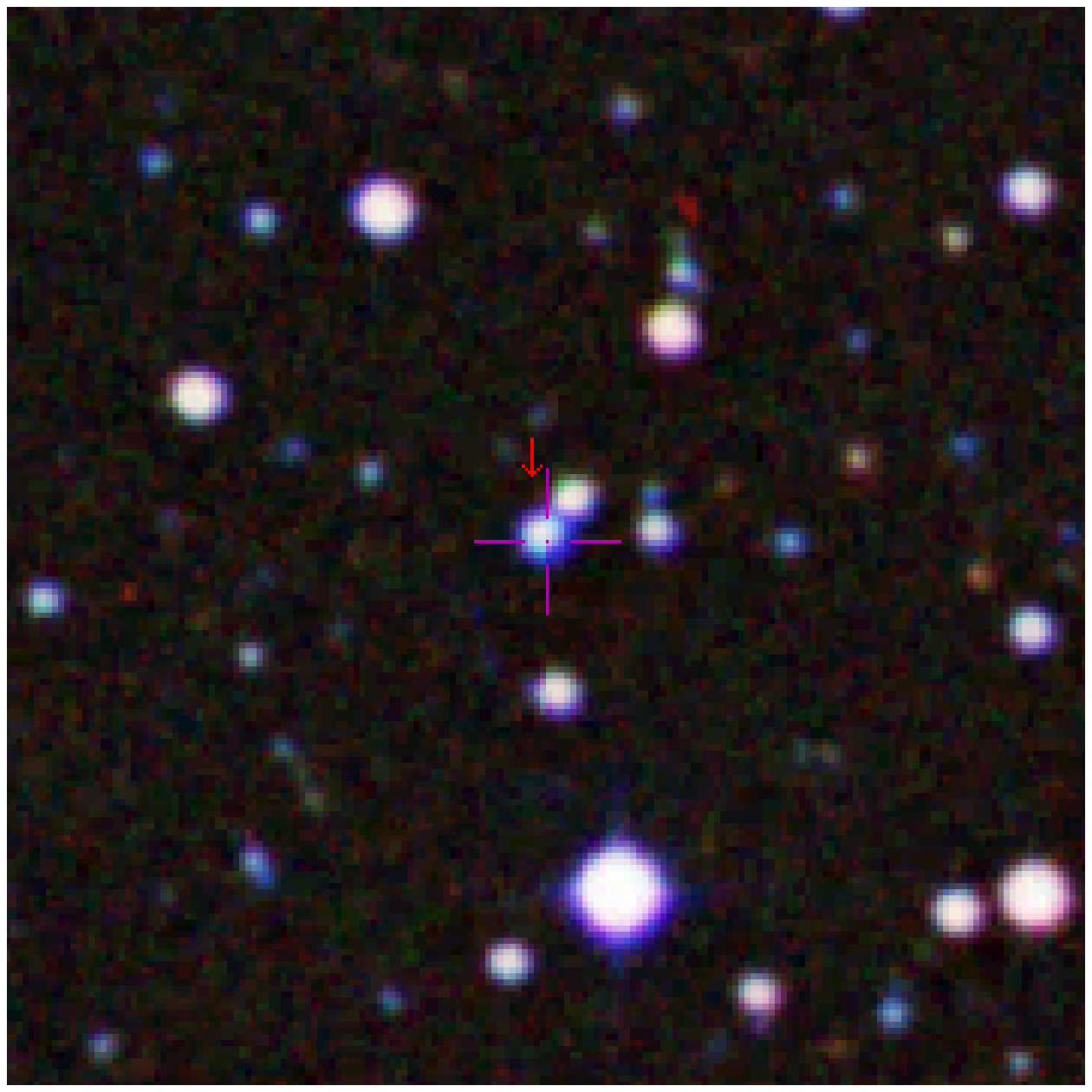}
\end{minipage}
\vspace{5mm}
\caption{Left: the field of KIC~6614501 in a real \kep\ image with the position
of the pixel array mask used during the Q7 run. The pixel size of about 3.98 
arcsec is not constant over time, differential velocity aberration results in 
the coordinates of the pixel mask changing constantly, typically by a few 
arcsec over each quarter. The three pixels actually used to form the optimal 
aperture, i.e. to collect photons from KIC~6614501, are those with a red dot
(colors available only in the electronic version).
Right: the same field in a SuperCosmos Sky Survey 3-color BRI image, which 
shows the target and its close companion at about 5.0 arcsec North West of it.
Both images are 2'$\times$2'.}
\label{fig:field}
\end{figure}

\section{Properties of the binary system}

\begin{figure}
\centering
\includegraphics[width=8.8cm]{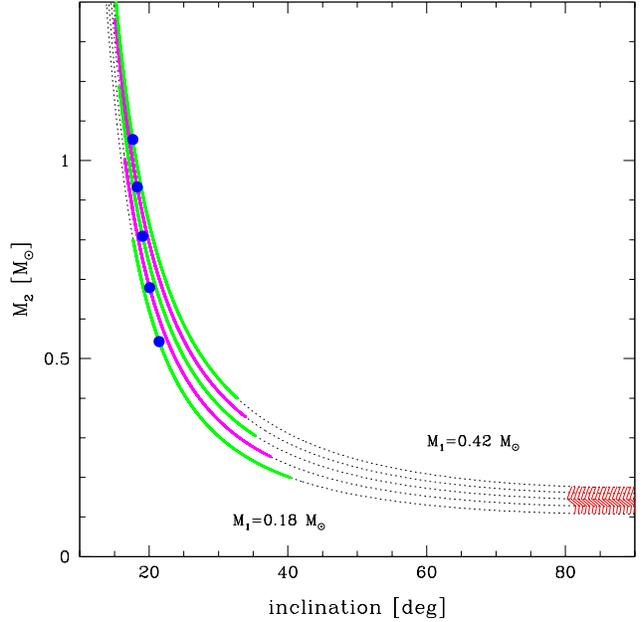}
\caption{Mass of the white dwarf secondary as a function of the system's 
inclination for a wide range of sdB masses, from 0.18 to 0.42 \msun\ with 0.06 
\msun\ steps.
The red region on the right is forbidden by the lack of eclipses.
When we consider also the amplitude of the ellipsoidal deformation of the sdB 
primary, we obtain a family of solutions represented by 
the blue dots and the green or violet 1$\sigma$ uncertainties 
(colors available only in the electronic version).
See text for more details.}
\label{fig:binary}
\end{figure}

The binary system is described by the mass function, that can be expressed
as follows:

\begin{equation}
{\rm sin}~i = K_1 ~ \left( \frac{P}{2 \pi G} \right)^{1/3} ~ \frac{(M_1 + 
M_2)}{M_2}^{2/3} 
\label{eq:mf}
\end{equation}

\noindent
In Fig.~6 the mass of the secondary~\footnote{From here on the terms 
``primary'' and ``secondary'' are used to indicate the brightest (sdB) 
and the faintest (WD) star of the binary, without any reference to the 
initially most massive star.} 
is shown as a function of the system's inclination for a wide range of 
sdB masses.
We see that the mass of the secondary is always larger than $\sim$0.1 \msun.

In order to constrain the inclination of the system and the mass of the 
secondary, we can consider also the ellipsoidal deformation of the primary.
The amplitude of the ellipsoidal modulation, equal to 
$368.0 \pm 11.7$
(section 4.1), is roughly proportional to 
$\approx (M_2/M_1) \ (R_1/a)^3 \ {\rm sin}^2 \ i$, 
where $a$ is the orbital semi-major axis and $R_1$ is the radius of the 
primary.
A more detailed treatment is given by \citet{morris93}.

The radius of the primary can be estimated from $R_1$~$=$~$(G M_1 / g)^{1/2}$, 
which gives for example $\sim$0.102 \rsun\ for $M_1$~$=$~$0.19$ \msun.
Using a gravity darkening coefficient of 0.479$\pm$0.010 (see section 5.1)
and adopting 0.27$\pm$0.02 for the linear limb darkening coefficient
in the {\em Kepler} passband \citep{claret11}, we can search for the best 
values of $i$ and $M_2$ that satisfy equation (3) of \citet{morris93}.
These solutions are shown in Fig.~6 as blue dots.
The uncertainties in green or violet (68\% confidence) were computed 
by means of the simpler equation (6) of \citealt[]{morris85}
(which, in our case, gives results almost identical to \citealt{morris93}, 
always within 2\%), taking into account the observational 
uncertainties in \logg, $K_1$, amplitude of the ellipsoidal deformation and 
orbital period.
For the mass of the primary we assumed an indetermination of $\pm 0.02$ \msun.
Despite the large uncertainties that we see in Fig.~6, the best values that 
we obtain for  $M_2$ are in good agreement with typical WD masses.
More importantly, Fig.~6 suggests a relatively low inclination $i\lsim 
40^{\circ}$, with a maximum probability near 20 degrees.
This is true for any primary mass and implies a minimum mass of 
$\sim$0.2 \msun\ for the secondary.
The large uncertainties on $i$ and $M_2$ in Fig.~6 are roughly 
proportional to the uncertainty in \logg, which dominates the error budget. 
A better determination of the surface gravity would be important to 
further constrain the system.

\subsection{Markov chain Monte Carlo simulations}

In order to confirm the analytical results of Fig.~6, and despite the level of 
degeneracy, we made an attempt to constrain the parameters 
of the system by modelling the light curve using the LCURVE light curve 
synthesis code and performing Markov chain Monte Carlo (MCMC) simulations 
(see \citealt{copperwheat10} and \citealt{bloemen11} for details of the code).
The gravity and limb darkening coefficients for the primary were computed 
using the same sdB atmosphere models as used in the computation of the beaming
factor (section 4.1),
assuming E(B--V)=0.06.
The gravity darkening coefficient was set to $0.479 \pm 0.010$, accounting for 
the uncertainty in \teff\ and \logg, while the limb darkening coefficient was 
obtained from a 4-parameter law (eq.~(5) of \citealt{claret11}).
Then we fixed the RV amplitude of the primary to the spectroscopic value, 
$K_1=97.2 \pm 2$ km/s, and we assumed a WD radius $R_2=0.015 \pm 0.003$ \rsun.
Actually this value has a much larger uncertainty (and there are no eclipses 
that constrain the radii) but this does not affect the results since the light 
contribution of the white dwarf is only $\sim$0.2\%.
Three MCMC runs were launched with a primary mass of 0.18, 0.24 and 0.36 \msun.
Inclination, mass ratio and beaming factor were kept as free parameters.
The results give an average inclination near 20 degrees, in agreement with the 
analytical results.
Moreover the formal uncertainties indicate an inclination below 50 degrees at a
95\% confidence level for all these primary masses.
However, these uncertainties are not very reliable because of convergence 
problems that are due to the very wide allowed range in the inclination and the
highly unlinear relation of the inclination with other parameters such as the 
mass ratio.
From the MCMC simulations we obtained a beaming factor of $1.33 \pm 0.03$,
confirming that the photometric contamination must be $\sim$12\%
in order to bring this value to $1.51 \pm 0.02$ as obtained from synthetic 
spectra.

\subsection{Different scenarios for the sdB evolution}

If the mass of KIC~6614501's primary is below $\sim$0.30 \msun, it must
be a He-core WD progenitor and we can compare its effective temperature and
surface gravity with He-core WD evolutionary tracks.
We obtain two possible scenarios:
i) for $M_1$ near 0.18-0.19 \msun, \teff\ and \logg\ are compatible with
a He-core white dwarf already on the final cooling track (after the possible 
H-shell flashes).
In this scenario a higher mass is excluded because the smaller radius
due to the electron degeneracy would be incompatible with the surface gravity
and with the amplitude of the ellipsoidal deformation.
Indeed, looking at \citet{panei07}, we see that there is only one model that,
having already reached the final cooling sequence, can fit \teff\
and \logg\ of KIC~6614501's primary: it is the model with a mass of 0.1869 
\msun\ between phase C and D, which corresponds to an age of about 180 Myr
from the end of the mass transfer.
The evolution is still relatively fast and the match between theoretical 
and observed \teff\ and \logg\ is valid only for about 1 Myr.
Masses smaller than $\sim$0.18 \msun\ (e.g. \citet{panei07}'s model with 0.1604
\msun) are ruled out because the star would never reach an effective 
temperature of 23~700~K.
ii) For $M_1\gsim0.19$ \msun, the sdB atmospheric parameters are compatible 
with a low-mass He-core WD progenitor that has not yet reached the final WD
cooling branch and is evolving along one of the cooling sequences related to
the H-shell flash episodes.
For example \teff\ and \logg\ are compatible with the 0.196 \msun\ (fourth 
cooling branch) or 0.242 \msun\ (first cooling branch) model of 
\citet{althaus01}.
The agreement between theoretical and observed \teff\ and \logg\ is valid for
less than 1 Myr or almost 2 Myr, and the age would be, respectively, 290 
or only 30 Myr.

If the mass of KIC~6614501's primary is larger than $\sim$0.32 \msun, there 
is another possibility:
iii) the sdB is a He-core burning object belonging to the extreme horizontal 
branch with a unusually low mass, between $\sim$0.32 and $\sim$0.40 \msun.
For such low masses the theoretical sdB evolutionary tracks \citep{han02}
pass through the position of KIC~6614501's primary in the \teff/\logg\ diagram.
To explore this possibility we computed a set of zero-age and terminal-age EHB 
(ZAEHB and TAEHB) for sdB masses of 0.33, 0.40 and 0.47 \msun, with different
envelope masses (0.00001, 0.0001, 0.001 and 0.005 \msun).
These ZAEHB and TAEHB tracks are shown in Fig.~7.
We see from Fig.~7 that, if the mass of KIC~6614501's primary was close 
to 0.33 \msun, the star would be close to the He-core exhaustion while, 
for a mass near 0.40 \msun, it would be close to the ZAEHB and much younger.
Such low-mass EHB stars are created from a progenitor with an initial mass 
between 2 and 2.5 \msun, for which the He-core ignition happens under 
non-degenerate conditions \citep{han02,hu07,pradamoroni09}.
Usually, sdB stars are considered to be post-He-flash objects and thus having 
a mass near 0.47 \msun.
However, binary population synthesis models by \citet{han03} show that 
sdB masses down to 0.32 \msun\ can occur in a close binary with a WD companion 
through the so-called second common envelope ejection channel.
Although these low sdB masses occur less frequently than the canonical 
0.47 \msun, they should not be neglected.

With the current data it is difficult to say which of the three scenarios 
proposed, He-core WD, He-core WD progenitor in a H-shell flash loop, 
or low-mass C/O-core WD progenitor, is the correct one.
We note, however, that the first hypothesis implies a secondary mass close to 
the peak of the WD mass distribution: for $M_1=0.19$ \msun, we obtain 
$i=21.2^{\circ} \pm ^{18.5}_{\hspace{1.3mm}3.7}$ and
$M_2=0.57 \pm ^{0.26}_{0.36}$ \msun.
An sdB mass greater by a factor of 2 would result in a higher luminosity 
by the same factor.
Thus, to know the distance of KIC~6614501 would help to discriminate between 
the different possibilities.
It is interesting to note that, from one hand, a low-mass EHB star would be 
compatible also with \teff\ and \logg\ of the other two objects close to 
KIC~6614501's primary in the \teff/\logg\ diagram (Fig.~7): HD~188112 and 
SDSS~J1625+3632.
But on the other hand, the known trigonometric distance of HD~188112 implies 
a mass of $0.24 \pm ^{0.10}_{0.07}$ \msun\ \citep{heber03}, which points 
towards a He-core WD progenitor, although a low-mass EHB of $\sim$0.32-0.34 
\msun\ can not be totally excluded.

In all the three scenarios proposed, KIC~6614501 is expected to merge 
in a time shorter than a Hubble time due to gravitational wave radiation. 
Following \citet{paczynski67}, the merging time varies between 0.9
(for $M_1$=0.40 and $M_2$=1.01 \msun) 
and 3.1 Gyr (for $M_1$=0.18 and $M_2$=0.54 \msun).

\section{KIC~6614501's primary in the context of He-core white dwarfs}

Low-mass white dwarfs have masses below $\sim$0.45 \msun\ and they are 
generally thought to contain inert helium cores.
However, a carbon/oxygen (C/O) core is not excluded for masses down to $\sim$0.33 \msun\ \citep{panei07,pradamoroni09}.
Below 0.3 \msun, helium burning is ruled out and the white dwarfs must 
have helium cores.
These objects are also called extremely low mass (ELM) white dwarfs.
At least for solar metallicities, ELM white dwarfs can form only from 
binary systems.
At higher metallicity it is less clear and the possible detection of single 
low-mass white dwarfs in NGC~6791, one of the oldest (\lsim 8 Gyr) and most 
metal-rich ([Fe/H] $\approx$ +0.4) open clusters of our Galaxy, has prompted
a heated debate \citep{origlia06,kalirai07}.
The evolution of He-core white dwarf models with low- or high-metallicity 
progenitors has been studied by \citet{serenelli02} and \citet{althaus09} 
respectively.

With a mass between $\sim$0.18 and $\sim$0.40 \msun, the primary star of 
KIC~6614501 is a precursor of a low-mass white dwarf.
If the mass is larger than about 0.32 \msun, it may evolve into a low-mass 
C/O-core white dwarf.
For lower masses it will become a He-core WD.
If the mass is near 0.18-0.19 \msun, it may have already reached the final
ELM WD cooling sequence.

The ELM white dwarfs (and their precursors)~\footnote{The difference between 
ELM WDs and their precursors is subtle.
In the next lines we will simply use the term ELM WDs for all of them,
according with recent articles.}
are a rare class of stars.
A list of these post-RGB low-mass objects from the literature is given in 
Table~4. Actually the list we found was longer but we decided to show only
those with a high reliability, for which the orbital period was measured 
through radial velocities and/or high precision photometry from space (\kep).
Moreover, we selected only those with \logg$<$7.0, which corresponds to
exclude those with a mass higher than 0.3 \msun\ \citep[see][Fig.~8]{kilic11a}.
Some new objects recently found by the ELM survey \citep{brown12}, which do
not meet our criteria, are not in Table~4.
Similarly, we have not included the primary star of SDSS~J1257+5428 
\citep{marsh11} because of the large uncertainties, in particular in \logg.
In most cases, the masses reported in Table~4 are obtained from evolutionary 
tracks and hence are model dependent.
For example, the evolutionary tracks of \citet{panei07} differ from those of
\citet{driebe98}, 
in particular for what concerns very low masses.
\citet{panei07} have shown that, when element diffusion is included, 
a dichotomy appears for stellar masses near 0.17 \msun: below this limit, 
the envelope is thick enough for residual nuclear burning that slows down 
the evolution.
For most systems in Table~4 that have a WD companion, only the minimum
mass of the secondary is known and thus a more massive neutron star or 
black hole companion can not be excluded.
The position of the objects of Table~4 in the \teff/\logg\ plane is shown
in Fig.~7.
In this figure we note that KIC~6614501 is the hottest object,
in a region of the \teff/\logg\ plane that is still quite empty.
Only other two stars reside in the same region: HD~188112, at slightly lower 
\teff, and SDSS~J1625+3632 at almost identical \teff\ but higher gravity.
These objects are crucial to study the limit between He-core burning EHB stars,
that will become low-mass C/O-core white dwarfs, and He-core WD progenitors.
They can be important also to study the complex evolutionary path that, through
the H-shell flashes, leads to the formation of He-core (ELM) white dwarfs.

\begin{figure}
\centering
\includegraphics[width=8.8cm]{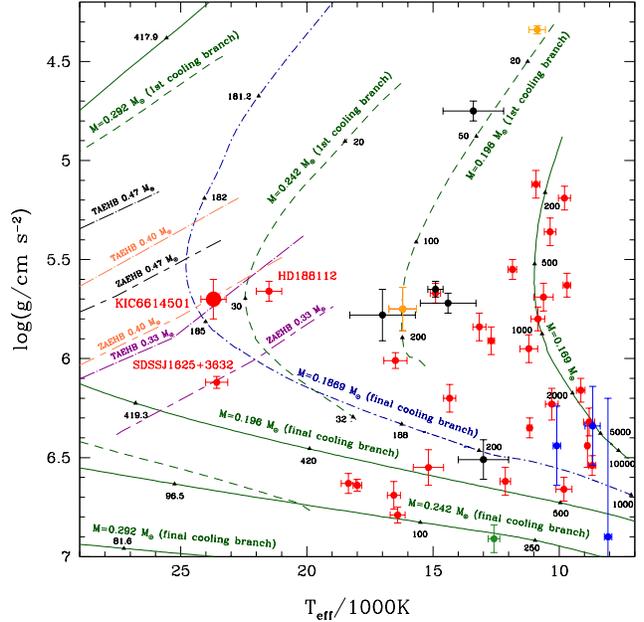}
\caption{ELM white dwarfs in the \teff/\logg\ plane.
These objects are in binary systems with a main sequence star, a white dwarf, 
or a neutron star (black, red, or blue symbols respectively, colors available
only in the electronic version).
The two objects in yellow are in globular clusters.
The green symbol corresponds to SDSS~J1448+1342, that could be either the 
component of a pole-on binary or a single star.
More details on the single objects are given in Table~4. 
The evolutionary tracks are from \citet{althaus01}, with a mass of 0.169, 
0.196, 0.242 and 0.292 \msun.
For the models with a mass greater than 0.17 \msun, which experience at least 
one thermonuclear H-shell flash, we show only the first and the last
cooling branch (those with a longer duration and a higher probability to be 
populated) for clarity.
Moreover we have included also the final cooling branch of the 0.1868 \msun\ 
model from \citet{panei07}, which matches very well the atmospheric parameters
of KIC~6614501's primary.
The numbers close to the triangles are the ages in Myrs from 
the end of the mass transfer.
The ZAEHB and TAEHB tracks for sdB masses of 0.33, 0.40 and 0.47 \msun\
are reported and correspond to a range of envelope masses between 0.00001 and 
0.005 \msun.}
\label{fig:binary}
\end{figure}

\begin{table*} 
\caption[]{ELM white dwarfs.}
\begin{tabular}{lccccccccc}
\hline
System name      & $g$ mag$^a$ & \teff/K & log($g$/cm~s$^{-2})$ & M$_1$/\msun & P$_{orb}$/d & Ecl.$^b$ & Comp. & M$_2$/\msun & Ref\\
\hline
%
%
V209~$\omega$~Cen&\hspace{1.5mm}V=16.58& 10866 $\pm$ 323& 4.34 $\pm$ 0.02 & 0.14 & 0.834 & 2 & reborn WD & \hspace{2.3mm}0.94 & 1\\
WASP~J0247$-$25  &\hspace{1.5mm}V=12.44 & \hspace{0.5mm}13400 $\pm$1200 & 4.75 $\pm$ 0.05 & 0.23 & 0.668 & 2 & A-type & \hspace{2.3mm}1.48 & 2\\
SDSS~J1233+1602  & 19.83 & 10920 $\pm$ 160 & 5.12 $\pm$ 0.07 & 0.17 & 0.151 &   & WD & $\ge$0.86 & 3 \\
SDSS~J1741+6526  & 18.27 & \hspace{1.4mm}9790 $\pm$ 240 & 5.19 $\pm$ 0.06 & 0.16 & 0.061 & & WD & $\ge$1.10 & 4 \\
SDSS~J2119$-$0018& 20.00 & 10360 $\pm$ 230 & 5.36 $\pm$ 0.07 & 0.17 & 0.087 &   & WD & $\ge$0.75 & 3 \\
SDSS~J0917+4638  & 18.70 & 11850 $\pm$ 170 & 5.55 $\pm$ 0.05 & 0.17 & 0.316 &   & WD & $\ge$0.27 & 3 \\
SDSS~J0112+1835  & 17.11 & \hspace{1.4mm}9690 $\pm$ 150 & 5.63 $\pm$ 0.06 & 0.16 & 0.147 &	& WD & $\ge$0.62 & 4 \\					
KIC~10657664     &\hspace{1.4mm}13.08$^c$ & \hspace{1.4mm}14900 $\pm$ 300$^d$ & \hspace{1.6mm}5.65 $\pm$ 0.04$^d$ & \hspace{1.5mm}0.37$^d$ & 3.273 & 2 & A-type & \hspace{0.9mm}2.5 & 5\\
HD~188112        & V=10.2& 21500 $\pm$ 500 & 5.66 $\pm$ 0.05 & 0.24 & 0.607 &   & WD & $\ge$0.73 & 6 \\
GALLEX~J1717     & V=13.7& 14900 $\pm$ 200 & 5.67 $\pm$ 0.05 & 0.18 & 0.246 & 1 & WD & \hspace{0.9mm}0.9 & 7 \\
SDSS~J0818+3536  & 20.48 & 10620 $\pm$ 380 & 5.69 $\pm$ 0.07 & 0.17 & 0.183 &   & WD & $\ge$0.26 & 3 \\
\hline
KIC~06614501     &\hspace{1.4mm}15.80$^c$ & 23700 $\pm$ 500 & 5.70 $\pm$ 0.10 & 0.24 & 0.157 & 0 & WD & $\sim$0.5-1.1 & 8 \\  
\hline
KOI~1224$^e$     &\hspace{1.4mm}14.01$^c$ & \hspace{0.5mm}14400 $\pm$1100 & 5.72 $\pm$ 0.05 & 0.20 & 2.698 & 2 & A/F & \hspace{2.3mm}1.59 & 9 \\
NGC~6121~V46     &V=18.5 & 16200 $\pm$ 550 & 5.75 $\pm$ 0.11 & 0.19 & 0.087 & 0 & WD & $\ge$0.26 & 10\\
KOI~81$^f$       &\hspace{1.4mm}11.30$^c$ & \hspace{0.5mm}17000 $\pm$1300 & 5.78 $\pm$ 0.13 & \hspace{-1.5mm}0.3 &\hspace{-1.2mm}23.878 & 2 & B9-type & \hspace{0.9mm}2.7 & 11 \\
SDSS~J0152+0749  & 18.01 & 10840 $\pm$ 270 & 5.80 $\pm$ 0.06 & 0.17 & 0.323 &	& WD & $\ge$0.57 & 4 \\  
SDSS~J0755+4906  & 20.09 & 13160 $\pm$ 260 & 5.84 $\pm$ 0.07 & 0.17 & 0.063 &   & WD & $\ge$0.81 & 3 \\
SDSS~J1422+4352  & 19.79 & 12690 $\pm$ 130 & 5.91 $\pm$ 0.07 & 0.17 & 0.379 &   & WD & $\ge$0.41 & 3 \\
SDSS~J1630+2712  & 20.04 & 11200 $\pm$ 350 & 5.95 $\pm$ 0.07 & 0.17 & 0.276 &   & WD & $\ge$0.52 & 3 \\
SDSS~J0106--1000 & \hspace{-1.4mm}19.8 & 16490 $\pm$ 460 & 6.01 $\pm$ 0.04 & 0.17 & 0.027 & 0 & WD & \hspace{0.9mm}0.4 & 12 \\
SDSS~J1625+3632  & 19.36 & 23570 $\pm$ 440 & 6.12 $\pm$ 0.03 & 0.20 & 0.232 &   & WD & $\ge$0.07 & 13 \\
SDSS~J1840+6423  & 18.76 & \hspace{1.4mm}9140 $\pm$ 170 & 6.16 $\pm$ 0.06 & 0.17 & 0.191 &   & WD & $\ge$0.64 & 4 \\
SDSS~J1439+1002  & 17.81 & 14340 $\pm$ 240 & 6.20 $\pm$ 0.07 & 0.18 & 0.437 &   & WD & $\ge$0.46 & 3 \\
SDSS~J0849+0445  & 19.31 & 10290 $\pm$ 250 & 6.23 $\pm$	0.08 & 0.17 & 0.079 &   & WD & $\ge$0.64 & 14 \\
SDSS~J1443+1509  & 18.58 &  \hspace{1.4mm}8810 $\pm$ 220 & 6.32 $\pm$	0.07 & 0.17 & 0.191 &   & WD & $\ge$0.83 & 4 \\
PSR~J1012+5307   & V=19.6& \hspace{1.4mm}8670 $\pm$ 300 & 6.34 $\pm$ 0.20 & 0.16 & 0.605 &  & PSR & \hspace{0.9mm}1.6 & 15,16 \\
LP~400-22        &\hspace{1.5mm}V=17.22& 11170 $\pm$ \hspace{1.4mm}90& 6.35 $\pm$ 0.05 & 0.19 & 1.010 &   & WD & $\ge$0.41 & 17,18,19 \\
PSR~J1911$-$5958 & V=22.1& 10090 $\pm$ 150 & 6.44 $\pm$ 0.20 & 0.18 & 0.837 &  & PSR & \hspace{2.3mm}1.34 & 20 \\
SDSS~J0822+2753  & 18.33 &  \hspace{1.4mm}8880 $\pm$ \hspace{1.4mm}60& 6.44 $\pm$ 0.11 & 0.17 & 0.244 &   & WD & $\ge$0.76 & 14 \\ 
KOI~74$^g$       & \hspace{1.4mm}10.85$^c$ & \hspace{0.5mm}13000 $\pm$1000 & 6.51 $\pm$ 0.10 & 0.23 & 5.189 & 2 & A1-type & \hspace{0.9mm}2.2 & 11,21 \\
NLTT~11748       &\hspace{1.4mm}V=16.7$^h$& \hspace{1.4mm}8690 $\pm$ 140 & 6.54 $\pm$ 0.05 & 0.18 & 0.235 & 2 & WD & \hspace{2.3mm}0.76 & 22,23,24,25,26\\
SDSS~J1053+5200  & 18.93 & 15180 $\pm$ 600 & 6.55 $\pm$ 0.09 & 0.20 & 0.043 &   & WD & $\ge$0.26 & 14,3 \\
SDSS~J1512+2615  & 19.24 & 12130 $\pm$ 210 & 6.62 $\pm$ 0.07 & 0.20 & 0.600 &   & WD & $\ge$0.28 & 3 \\
SDSS~J0923+3028  & 15.63 & 18350 $\pm$ 290 & 6.63 $\pm$ 0.05 & 0.23 & 0.045 &   & WD & $\ge$0.34 & 3 \\
SDSS~J1234$-$0228& 17.86 & \hspace{1.1mm}18000 $\pm$ 170$^i$ & \hspace{1.4mm}6.64 $\pm$ 0.03$^i$ & 0.23 & 0.091 &   & WD & $\ge$0.09 & 13 \\ 
SDSS~J1518+0658  & 17.46 &  \hspace{1.4mm}9810 $\pm$ 320 & 6.66 $\pm$	0.06 & 0.20 & 0.609 &	& WD & $\ge$0.58 & 4 \\
SDSS~J1436+5010  & 18.23 & 16550 $\pm$ 260 & 6.69 $\pm$	0.07 & 0.24 & 0.046 &   & WD & $\ge$0.46 & 14 \\
SDSS~J0651+2844  & \hspace{-1.4mm}19.1 & 16400 $\pm$ 300 & 6.79 $\pm$	0.04 & 0.25 & 0.009 & 2 & WD & \hspace{2.3mm}0.55 & 27 \\
PSR~J0218+4232   & V=24.2& \hspace{1.4mm}8060 $\pm$ 150 & 6.9  $\pm$ 0.7  & \hspace{-1.5mm}0.2 & 2.029 &  & PSR & \hspace{0.9mm}1.6 & 28\\
SDSS~J1448+1342  & 19.22 & 12580 $\pm$ 230 & 6.91 $\pm$ 0.07 & 0.25 & $^l$  &   &    &  & 3 \\
%
%
\hline

\multicolumn{10}{l}{Notes: $^a$ system's magnitude; $^b$ eclipse: 0=not detected, 1=secondary ecl. detected, 2=secondary and primary ecl. detected; $^c$ from Kepler Input}\\
\multicolumn{10}{l}{\hspace{7.2mm} Catalogue; $^d$ a secondary solution from the same authors gives \teff/K=14600~$\pm$~300, \logg=5.50~$\pm$~0.02, M/\msun=0.26; $^e$ KIC~6606653;}\\
\multicolumn{10}{l}{\hspace{7.2mm} $^f$ KIC~8823868; $^g$ KIC~6889235; $^h$ from USNO; $^i$ \teff/K=17470~$\pm$~750, \logg=6.38~$\pm$~0.05 from \citealt{liebert04}; $^l$ can be either a}\\
\multicolumn{10}{l}{\hspace{7.2mm} pole-on binary or a single ELM WD, see \citealt{brown10} for more details.}\\
\multicolumn{10}{l}{References: 1 \citealt{kaluzny07}, 2 \citealt{maxted11}, 3 \citealt{brown10}, 4 \citealt{brown12}, 5 \citealt{carter11}, 6 \citealt{heber03},}\\
\multicolumn{10}{l}{\hspace{13.3mm} 7 \citealt{vennes11}, 8 this paper, 9 \citealt{breton12}, 10 \citealt{o'toole06}, 11 \citealt{vankerkwijk10}, 12 \citealt{kilic11b},}\\
\multicolumn{10}{l}{\hspace{13.3mm} 13 \citealt{kilic11a}, 14 \citealt{kilic10a}, 15 \citealt{vankerkwijk96}, 16 \citealt{callanan98}, 17 \citealt{vennes09}, 18 \citealt{kilic09},}\\
\multicolumn{10}{l}{\hspace{13.3mm} 19 \citealt{kawka06}, 20 \citealt{bassa06}, 21 \citealt{bloemen12}, 22 \citealt{kawka09}, 23 \citealt{kawka10},}\\
\multicolumn{10}{l}{\hspace{13.3mm} 24 \citealt{steinfadt10}, 25 \citealt{kilic10b}, 26 \citealt{shporer10}, 27 \citealt{brown11}, 28 \citealt{bassa03}.}\\
\multicolumn{10}{l}{\hspace{13.3mm}}\\
%
%
\end{tabular}
\end{table*}

\section*{Acknowledgments}

The authors thank Conny Aerts and Andrew Tkachenko for the work done in 
organising and leading the KASC WG9 on binary stars and maintening its web
site, making easier the realisation of this article.
They also thank Ronald L.~Gilliland, Martin Still and Karen Kinemuchi for
useful informations on the \kep\ pixel data;
and the reviewer, Stephan Vennes, for helpful comments on the paper.
R.S. was supported by the PRIN-INAF on ``Asteroseismology: looking inside
the stars with space- and ground-based observations''.
R.H.\O. and S.B. have received support through the European Research Council 
under the European Community's Seventh Framework Programme
(FP7/2007--2013)/ERC grant agreement N$^{\underline{\mathrm o}}$\,227224
({\sc prosperity}), as well as from the Research Council of K.U. Leuven grant
agreement GOA/2008/04.
M.D.R. and L.E.F. were supported by the Missouri Space Grant, funded by NASA.
P.D. is a Postdoctoral Fellow of the Fund for Scientific Research 
of Flanders (FWO).
H.H. was supported by the Netherlands Organisation for Scientific Research 
(NWO).
T.R.M. was supported by the Science and Technology Facilities Council.
Finally, the authors gratefully acknowledge the \kep\ team and everybody who
has contributed to making this mission possible.
Funding for the \kepmi\ is provided by NASA's Science Mission Directorate.

\bibliographystyle{mn2e}
\bibliography{sdbrefs}

\label{lastpage}

\end{document}